\documentclass[10pt,amsmaths,preprinnumbers]{article}
\usepackage{psfig}
\usepackage{graphicx}
\usepackage{epsfig}
\usepackage{latexsym}
\usepackage{bm}
\newcommand{\be}{\begin{equation}}
\newcommand{\ee}{\end{equation}}
\newcommand{\ben}{\begin{eqnarray}}
\newcommand{\een}{\end{eqnarray}}
\newcommand{\bF}{\begin{figure}}
\newcommand{\eF}{\end{figure}}
\begin{document}
\title{\bf Klein Paradox in Bosons}
\author{Partha Ghose and Manoj K. Samal\\
{\it\small S. N. Bose National Centre for Basic Sciences,\newline
  Block JD, Sector III, Salt Lake, Kolkata 700 098, India}\and
Animesh Datta \footnote{Corresponding author: animesh.datta@iitk.ac.in
(Animesh Datta)}\\
{\it\small Department of Electrical Engineering, Indian Institute
of Technology, Kanpur, 208 016, India}}
\maketitle
\begin{abstract}
We show analytically that the {\it Zitterbewegung} and Klein Paradox, such well known aspects of the Dirac Equation are not found in the case of Bosons. We use the Kemmer-Duffin-Harish Chandra formalism with $\beta$ matrices to arrive at our results. 
\end{abstract}

\section{Introduction}
In discussing problems and interactions in which the electron (or any
fermion) is "spread out" over distances large compared to its Compton
wavelength, we may simply ignore the existence of the uninterpreted
negative-energy solutions of the one-particle relativistic Dirac
Equation \cite{dirac} and hope of obtaining physically sensible and
accurate results. This will not work, however, in situations which
find electrons (or fermions) localized to distances comparable with
$\frac{\hbar}{mc}$. The negative frequency amplitudes will then be
appreciable, {\it Zitterbewegung} \cite{erwin} terms in the current
important and indeed we shall find ourselves beset by paradoxes and
dilemmas which defy interpretation within the framework of the one
particle Dirac equation. A celebrated example of these difficulties is
the {\bf Klein Paradox} \cite{klein} for electrons, and in general for
fermions. It is discussed in detail in numerous texts
\cite{drell},\cite{greiner} and resolved using the hole-theory which
provides the treatment of the negative energy states. The actual
paradox is that in the situations under question, the coefficient of
reflection for an electron beam or any current from a barrier is more
than one. It is to be noted that all the energies under consideration
are below the threshold of pair creation.

In this Letter we address the same problem for bosons. We explicitly
show that the reflection coefficient is less than one and hence the
scope of any paradox is nonexistant.

The plan of the Letter is as follows. In Section 2 we explain the
formalism in some detail for the massive and massless bosons.
Section 3 contains our results about the absence of Klein Paradix like
phenomenon in the case of massive bosons. The corresponding result for
massless bosons is presented in Section 4. We make some concluding remarks in
Section 5.

\section{Formalism}
In order to address the problem of Klein Paradox, a single-particle
relativistic quantum mechanics is necessary. For fermions, such a
framework is provided by the Dirac Equation, though it eventually
turns out to be a multiparticle equation. Such a consistent relativistic quantum
mechanics for spin 0 and spin 1 bosons was provided by one of us \cite{pg}.
It has been shown in Ref \cite{pg} that a conserved four-current with a positive definite time component does exist for relativistic bosons, and is associated not with charge current but the flow of energy.
\subsection{The Kemmer Formalism for massive bosons}
       The Kemmer equations for a massive spin 0 or spin 1 boson are
\ben
(i\beta^{\mu}\partial_{\mu}+m)\psi&=&0 \\
i\partial^{\mu}\bar{\psi}\beta_{\mu}-m\bar{\psi}&=&0
\een

where the Kemmer-Duffin-Petiau matrices satisfy the algebra
\be
\label{algebra}
\beta_{\mu}\beta_{\nu}\beta_{\lambda}+\beta_{\lambda}\beta_{\nu}\beta_{\mu}=\beta_{\mu}g_{\nu\lambda}+\beta_{\lambda}g_{\nu\mu}
\ee
and 
$\bar{\psi}=\psi^{\dagger}\eta_0$ with $\eta_0=2\beta_0^2 -1$. As a
consequence of Eqn (\ref{algebra}) one has  
\be
\eta_0\beta_0=\beta_0\eta_0=\beta_0, \hspace{1cm} \eta_0^2=1
\ee
\be
\eta_0\beta_i+\beta_i\eta_0=0, \hspace{1cm} i= 1,2,3.
\ee

We choose a representation in which $\beta_i^{\dagger}=-\beta_i
 (i=1,2,3)$. The $\beta_i$ in Ref \cite{kemmer} correspond to what we
call $i\beta_i$ and $\beta_0^{\dagger}=\beta_0$. There are two
nontrivial representations of these $\beta$'s: one is $5\times5$,
which describe spin 0 bosons, the other is $10\times10$, which
describe spin 1 bosons. For their explicit forms the reader is
referred to \cite{lorentzpg}.

Multiplying (1) by $\partial^{\rho}\beta_{\rho}\beta_{\nu}$, one can
deduce, 
\be
\label{six}
\partial_{\nu}\psi = \partial^{\rho}\beta_{\rho}\beta_{\nu}\psi
\ee
and similarly
\be
\partial_{\nu}\bar{\psi} = \partial^{\rho}\bar{\psi}\beta_{\nu}\beta_{\rho}
\ee

The second order wave equation
\ben
(\Box + m^2)\psi&=&0 \\
(\Box + m^2)\bar{\psi}&=&0
\een
 follow from (1) and (\ref{six}) and (2) and (7) resepectively. A
 further cosequence of (1) and (2) is the conservation equation 
\ben
\partial_{\mu}j^{\mu}=0 \\
j^{\mu}= \bar{\psi}\beta^{\mu}\psi
\een 

Unfortunately, although $j_{\mu}$ is a conserved 4-vector, $j_0$ is
not positive definite and cannot be interpreted as probability
density, a feature common with the Klein- Gordon and Proca equations.

\subsection{Massless Bosons}

Harish Chandra \cite{hr} showed how the Kemmer formalism could be
applied for the case of massless bosons {\it without taking the limit
  $m \to 0$}. We restrict considerations to neutral particles
only. Instead of (1), one writes 
\be
\label{nomass}
i\beta_{\mu}\partial^{\mu}\psi+m\gamma\psi = 0
\ee 
where $\gamma$ is a matrix satisfying 
\ben
\gamma^2=\gamma \\
\gamma\beta_{\mu}+\beta_{\mu}\gamma=\beta_{\mu}
\een

Multiplying  (\ref{nomass}) by $(1-\gamma)$ on the left, one obtains 

\be
i\beta_{\mu}\partial^{\mu}(\gamma\psi)=0
\ee

Multiplying  (\ref{nomass}) by
$\partial_{\lambda}\beta^{\lambda}\beta^{\nu}$ on the left, one also
gets
\be
\partial^{\lambda}\beta_{\lambda}\beta_{\nu}(\gamma\psi)=\partial_{\nu}(\gamma\psi)
\ee

The second order wave equation 
\be
\Box(\gamma\psi)=0
\ee
follows from  (15) and (16) which show that $\gamma\psi$ is a massless boson.

\subsection{Probability Interpretation}
 Although $j_0$ is not positive definite, the fact that a conserved
 four- vector current with a positive definite time component can be
 defined using this formalism as follows. Multiplying (1) by
 $\beta_0$, one obtains the Schr\"odinger form of the equation,
\be
\label{sch}
i\frac{\partial\psi}{\partial t}=[-i\tilde\beta_i\partial_i-m\beta_0]\psi
\ee
where  $\tilde\beta_i=\beta_0\beta_i-\beta_i\beta_0$. Multiplying (1) by $(1-\beta_0^2)$ one obtains
\be
i\beta_i\beta_
0^2\partial_i\psi=-m(1-\beta_0^2)\psi.
\ee

If one multiplies Eqn (\ref{sch}) by $\psi^{\dagger}$ from the left,
  its Hermitian conjugate by $\psi$ from the right and adds the
  equations, one obtains the continuity equation
\be
\frac{\partial\psi^{\dagger}\psi}{\partial t} + \partial_i \psi^{\dagger}\tilde\beta_i\psi=0
\ee
whence $S_i=\psi^{\dagger}\tilde\beta_i\psi$ is the conserved
  four-vector current density in
  the direction $i$ with $S_0=\psi^{\dagger}\psi>0$.
For details the reader is referred to Ref \cite{saikat}, Sec 1.
\section{Massive Bosons}
In this section we present our results. We always consider a potential
step at $x=0$ with the incident wave travelling along the positive $x$ axis.
\subsection{Spin 1 Bosons}
The Kemmer-Duffin-HarishChandra wave function in this case is given by 
\be
\Psi = \frac{A}{\sqrt{m}}\left(\begin{array}{cc}
                                      \psi_1 \\
				      \psi_2 \\
				      \vdots \\ 
                                       \psi_9\\       
                                       \psi_{10} \end{array}\right)
\ee

 We consider the situation of a wave propagating along the positive x-axis with a potential step at x=0,
\ben
\Psi_{inc}&=& \frac{A}{\sqrt{m}}(0,0,E_0,0,-E_0,0,0,0,0,0)^T e^{i(kx- \omega t)} \\
\Psi_{ref}&=& \frac{B}{\sqrt{m}}(0,0,E_0,0,E_0,0,0,0,0,0)^T e^{-i(kx+ \omega t)} \\
\Psi_{trans}&=& \frac{C}{\sqrt{m}}(0,0,E_0,0,-\sqrt{\epsilon}E_0,0,0,0,0,0)^T e^{i(k'x-\omega t)}.
\een
where $\epsilon$ contains the information about the height of the
barrier. The exact relation is not essential to the context of the
problem.

 Then $\frac{S_{inc}}{S_{ref}}=\frac{\Psi^{\dagger}_{inc}{\tilde \beta_i}\Psi_{inc}}{\Psi^{\dagger}_{ref}{\tilde \beta_i}\Psi_{ref}}$ for $i = 1,2,3.$
 In particular case that we treat we have $i=1$. Then some algebra gives $\frac{S_{inc}}{S_{ref}}= \frac{|B|^2}{|A|^2}.$

Matching $\Psi_{inc}+\Psi_{ref}=\Psi_{trans}$ at $x=0, t=0$, readily
gives 
\ben 
A+B&=&C \\
A-B&=&\sqrt{\epsilon}C
\een
 whence $\frac{|B|^2}{|A|^2}=\Big(\frac{1-\sqrt{\epsilon}}{1+\sqrt{\epsilon}}\Big)^2$.  Thus the ratio $\frac{S_{ref}}{S_{inc}}< 1$ which means that no effect analogous to Klein paradox occurs in the case of Spin 1 massive bosons.

\subsection{Spin 0 Bosons}

For a wave propagating along $i=1$, with $x_0=t$, we have 
\ben
\label{mass1}
\Psi_{inc}&=& A (k,0,0,-ik_0,1) e^{i(k_0x_0- kx)} \\ \nonumber
\Psi_{ref}&=& B (-k,0,0,-ik_0,1) e^{-i(k_0x_0+ kx)} \\ \nonumber
\Psi_{trans}&=& C (k',0,0,-ik'_0,1) e^{i(k'_0x_0-k'x)}
\label{star}
\een
 
which are special cases of
\be
\Psi=\left(\begin{array}{cc}
                          \phi_1 \\
			  \phi_2 \\
			  \phi_3\\       
                           \phi_{0}\\
                           \phi \end{array}\right)
\ee
 where $\phi_i=\partial_i \phi$ and $\phi=e^{ik_{\mu}x_{\mu}}=e^{i(k_0x_0-k\cdot x)}.$
The barrier is as usual at $x=0$ and the information about $x>0$ is
incorporated in $k'$ and $k'_0$. 
Since $S_i= \Psi^{\dagger}\beta_i\Psi$ for all $i$ and 
\be
\beta_1=\left(\begin{array}{cccccc}
			 0 &0 & 0 &-i  & 0  \\
			 0 &0 & 0 & 0  & 0 \\ 
			 0 &0 & 0 & 0  & 0  \\
			 i &0 & 0 & 0  & 0  \\
			 0 &0 & 0 & 0  & 0 \end{array}\right)
\ee
 $\frac{S_{ref}}{S_{inc}}=\frac{|B|^2}{|A|^2} = \big|\frac{B}{A}\big|^2.$

The set of Eqns (\ref{mass1}) evaluated at $x=0,x_0=0$ give
\ben
\label{three}
A - B &=& C (k'/k) \\ \nonumber
A + B &=& C (k'_0/k_0) \\ \nonumber
A + B &=& C
\een
Now $E^2=\hbar^2 c^2(k^2_0-k^2)+m^2_0c^4 $ for $x<0$ and  $E^2=\hbar^2
c^2(k'^2_0-k'^2)+m^2_0c^4 $ for $x>0$. If the frequency of the wave
does not change on moving into the potential {\it i.e.,}changing the media, which is indeed the case, then $k'_0=k_0$ and consequently $k'^2>k^2.$ 

Finally, Eqn (\ref{three}) gives $|\frac{B}{A}|=\big(\frac{1-k'/k}{1+k'/k}\big)^2.$

Hence there is no evidence of Klein Paradox in the system under question as the reflection coefficient is less than 1.

\section{Massless Bosons}
In this case we will only be considering neutral bosons of spin 1,
{\it i.e,} photons. 
Using the Harish Chandra formalism \cite{hr}, we have for a beam of photons propagating along the positive  $x$ axis, 
\be
\gamma\Psi=\left(\begin{array}{cc}
                          0 \\
			  0 \\
			  E_z\\       
                           0\\
                           H_y \\
			   0 \\
			   0 \\
			   0 \\
			   0 \\
			   0\end{array}\right)
\ee
 The barrier is again placed at $x=0$. 
Let 
\ben
\label{four}
E^i_z &=& E_0 e^{i(kx - \omega t - \phi)}\\ \nonumber
E^r_z &=& E_0 \sqrt{R}e^{-i(kx - \omega t + \phi)} \\ \nonumber
E^t_z &=& E_0 \sqrt{T}e^{i(k'x - \omega t + \chi)}.
\een
 Maxwell's Equations then at once give (using $\partial_x E_z=\partial_tH_y$)
\ben
\label{five}
H^i_y &=& -\frac{k}{\omega} E_0 e^{i(kx - \omega t - \phi)} \\ \nonumber
H^r_y &=& -\frac{k}{\omega} E_0 \sqrt{R}e^{-i(kx + \omega t + \phi)} \\ \nonumber
H^t_y &=& -\frac{k}{\omega} E_0 \sqrt{T}e^{i(k'x - \omega t - \chi)}.
\een

Now, $S_i=[E \times H]_i$ is the (Poynting vector) current density as given by \cite{lorentzpg}.
Thus from Eqns (\ref{four}) and (\ref{five})
\be
\frac{S^{ref}_x}{S^{inc}_x} = R.
\ee

Matching $E^i_z - E^r_z = E^t_z$ at $x=0,t=0$ gives at once
\be
R = \frac{S^{ref}_x}{S^{inc}_x}=\Big(\frac{k'/k - 1}{k'/k + 1}\Big)^2 < 1 .
\ee

Thus there is no signature of Klein Paradox in the case of photons. 

\section{Conclusion}

The Klein Paradox was a generic problem for the Dirac equation that
was resolved very soon after its inception. The Klein- Gordon equation also shows the paradox \cite{baym} where it is an artefact of its not being a proper quantum mechanical equation with a positive definite probability density. However, to the best of
our knowledge the question has never been asked, lest answered for
bosons. This Letter asks that natural question and conclusively
answers it by proving that the Klein paradox does not exist in the case of bosons. The
paradox for fermions was resolved using the Hole theory by Dirac. Here
too, the relativistic equations for spin 0 and spin 1 bosons are used. As has been shown in Section
2, the Klein-Gordon equation which is generally used to describe spin
0 particles can be deduced from the basic equation of
Kemmer. For a similar deduction of Maxwell's equations the reader is
referred to Ref \cite{pg}, Sec 5. Our results, based on sound
  mathematical footing answer a basic question. However, there have
  been recent results based on Bohmian trajectories \cite{inn} showing
  the absence of Klein paradox in the case of outgoing scattering asymptotics.

\section*{Acknowledgements}
P. G. acknowledges financial support from DST, Government of
India. A. D. thanks the S. N. Bose National Centre for Basic Sciences
for hospitality that enabled this work to be undertaken.

The authors also thank the reviewers whose comments helped improve the
clarity of presentation.


\begin{thebibliography}{99}
\bibitem{dirac} P. A. M. Dirac, Proc. Roy. Soc. (London), A 117, 610 (1928), {\it ibid} A 118, 351 (1928). "Principles of Quantum Mechanics", {\it op. cit}.
\bibitem{erwin} E. Schrodinger, Sitzber. Prevss. Akad. Wiss. Physik-Math, 24, 418 (1930).
\bibitem{klein} O. Klein, Z. Physik, 53, 157 (1929).
\bibitem{drell} J. D. Bjorken, S. D. Drell, Relativistic Quantum Mechanics, McGraw Hill, New York, 1965.
\bibitem{greiner} W. Greiner, Relativistic Quantum Mechanics, Springer- Verlag, Berlin, 1997.
\bibitem{pg} P. Ghose, Found. of Physics, 26 (11), 1441, 1996.
\bibitem{kemmer} N. Kemmer, Proc. Roy. Soc A, 173, 91, 1939.
\bibitem{lorentzpg} P. Ghose, M. K. Samal, Phys. Rev. E, 036620, 2001.
\bibitem{hr} Harish Chandra, Proc. Roy. Soc(London), 186, 502, 1946.
\bibitem{saikat} P. Ghose, A. S. Majumdar, S. Guha, J. Sau,
  Phys. Lett. A, 290 (2001), 205.
\bibitem{inn} G. Grubl {\it et al}, J. Phys. A: Math, Gen, 34, 2753- 2764 (2001).
\bibitem{baym} G. Baym, Lectures of Quantum Mechanics, W. A. Benjamin. Inc., New York, 1969. 
\end{thebibliography}
\end{document}